\documentclass[a4paper,12pt,pre]{revtex4-1}
\usepackage{graphicx}
\usepackage{amsmath}
\usepackage{fancyhdr}
\usepackage{color}
\usepackage{amssymb}
\setcitestyle{round}
\begin{document}
\title{Quantifying extrinsic noise in gene expression using the maximum entropy framework}
\author{Purushottam D. Dixit}
\affiliation{Biosciences Department, Brookhaven National Laboratory, Upton NY 11973}
\thanks{Corresponding author: Phone: (631) 344-3742; Email: pdixit@bnl.gov}

\begin{abstract}
\end{abstract}
\maketitle
\section*{Abstract}
We present a maximum entropy framework to separate  intrinsic and  extrinsic contributions to noisy gene expression solely from the profile of  expression. We express the experimentally accessible probability distribution of the copy number of the gene product (mRNA or protein) by accounting for possible variations in extrinsic factors. The distribution of extrinsic factors is estimated using the maximum entropy principle. Our results show that extrinsic factors  qualitatively and quantitatively affect the  probability distribution of the gene product.  We work out, in detail, the transcription of mRNA from a constitutively expressed promoter in {\it E. coli}. We suggest that the variation in extrinsic factors may account for the observed {\it  wider than Poisson} distribution of mRNA copy numbers. We successfully test our framework on a numerical simulation of a simple gene expression scheme that accounts for the variation in extrinsic factors. We also make falsifiable predictions, some of which are tested on previous experiments in {\it E. coli} while others need verification. Application of the current framework to more complex situations is also discussed.

\vspace{5mm}

{\it Keywords: Stochastic gene expression, maximum entropy}

\pagebreak
\section{Introduction}

Recent experiments show that the life cycle of a gene product inside the cell is stochastic. For any gene, there exists great cell to cell variation in the expression level of both the protein and the mRNA~\citep{Bar-Even2006,Cai2006,Elowitz2007,Kaufmann2007,Raj2008,Rosenfeld2005,Taniguchi2010,Newman2006,Paulsson2005,Raj2006} and changing this variation has phenotypical and fitness effects~\citep{Kaern2005,Maheshri2007,mammar2007,Fraser2004}. Recently, it was also shown that co-regulated proteins have correlated variability~\citep{Stewart-Ornstein2012}.  This variation arises from a) the `intrinsic' statistical mechanical fluctuations in diffusion and binding of the molecules involved in gene expression and b) the variation in  `extrinsic' factors that determine the state of the cell. Examples of extrinsic factors include the external environment~\citep{chubb2006transcriptional,golding2006eukaryotic}, the epigenetic state of the cell~\citep{so2011general,golding2005real}, the time from last cell division, levels of molecular machines such as RNA polymerase, ribosome, proteases, and RNAses~\citep{Elowitz2007,Kaufmann2007,Swain2002}.  In a given population of cells, the total noise (coefficient of variation)
\begin{eqnarray}\eta_{\rm T} &=& \frac{\langle {\rm m}^2 \rangle - \langle {\rm m} \rangle^2}{\langle {\rm m} \rangle^2}\end{eqnarray} 
serves as a useful experimental quantification of the variability in gene expression where $\langle {\rm m}\rangle$ is the mean level of  the gene product ${\rm m}$ (mRNA or protein) and $\langle {\rm m^2} \rangle -\langle {\rm m}\rangle^2$ is the variance.

For a constitutively expressing promoter, under simplifying conditions, the contribution to  $\eta_{\rm T}$ associated with extrinsic factors, the extrinsic noise $\eta_{\rm E}$, can be experimentally measured separately from the intrinsic noise $\eta_{\rm I}$~\citep{Swain2002,Elowitz2007,Rosenfeld2005,Stewart-Ornstein2012}. It is now known that the extrinsic noise is the dominant contributor to gene expression~\citep{Elowitz2007,Stewart-Ornstein2012} and can change the profile of gene expression in a non-trivial manner~\citep{shahrezaei2008colored}. Evidently, an important step towards the conceptual understanding of the noisy gene expression is to quantitatively account for the effect of variations in extrinsic factors on gene expression.

The major technical hurdle in building a comprehensive theory for extrinsic variation originates in the multitude of factors that contribute to it. Consequently, theoretical exploration of noisy gene expression has concentrated on intrinsic noise. Here, one generally employs the master equation framework~\citep{Thattai2001,Raj2006,Sanchez2008,Hemberg2007,Paulsson2005}. Briefly, we define a set of reactions $\mathcal R$ involving species $\mathcal G$ (protein, mRNA, etc.). A transition matrix for evolution of the probability distribution of  $\mathcal G$ is constructed. The transition matrix contains information about the chemistry (rates, allosteric binding, etc) and the topology (feedback, loops, etc) of the reactions. The probability distribution $P(\mathcal G | t,  \mathcal K)$  is then sought in terms of the rate constants $\mathcal K = \{k_1, k_2, \dots \}$ of all reactions and time $t$. Since closed form solutions for the master equation exist only for a few simple systems, much theoretical development explores efficient ways of simplifying the solution of the master equation~\citep{Raj2006,Sanchez2008,friedman2006linking}.

The chemical reactions are carried out by molecular machines e.g. RNA polymerase, ribosomes, enzymes, etc.  Moreover, these chemical reactions also depend on the chemical state of the cell for example the time from cell division, the chromatin structure of DNA, presence of DNA binding proteins, RNA degradation by small RNAs, presence of RNA binding proteins, etc. All these variables differ from cell to cell and as a function of time. Hence, $\mathcal K$ depend on the state of the cell and are themselves stochastic variables. This  makes gene expression a doubly stochastic process~\citep{dixit2012,dsp}. We interpret the variability in $\mathcal K$ --- which represents the variability in global factors --- as the extrinsic variability. 

Due to the very large number of affectors, it is impossible to model the extrinsic variability from first principles.  Consequently, the theoretical treatment has either assumed a small extrinsic contribution resulting in a linear susceptibility-like analysis~\citep{Swain2002} or assumed an {\it ad hoc} structure for the distribution of extrinsic factors~\citep{shahrezaei2008colored,kaern:chaos2006}.  Here, instead of accounting for {\it all} the extrinsic contributors {\it ab initio}, we develop a maximum entropy framework to estimate $P(\mathcal K)$, from limited information about the gene expression profile.  We successfully test our results on a simplified numerical scheme for mRNA production that explicitly incorporates the variability in molecular machinery. Most importantly, we show that extrinsic factors can   qualitatively and quantitatively affect the experimentally observed histogram of the gene expression product (protein or mRNA).

%
\section{Theory}



\begin{figure}[h]
	\includegraphics[scale=0.2]{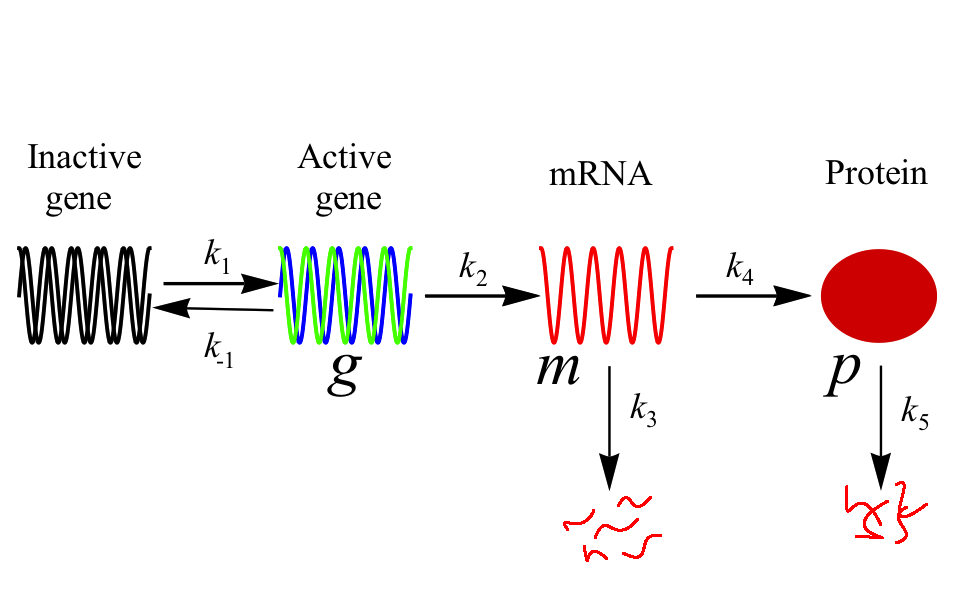}
	\caption{\label{fg:constitutive}}
\end{figure}

For concreteness, consider a contitutively expressing promoter in a bacterial setting (see Fig.~\ref{fg:constitutive}). Later, we will substantially simplify this example. Here, an inactive gene is converted to an active gene with rate constant $k_1$ and vice versa (rate constant $k_{-1}$). An mRNA molecule is transcribed from the active gene at a rate constant $k_2$. A protein is translated from the mRNA at a rate $k_4$. The mRNA and the protein are degraded at rates $k_3$ and $k_5$ respectively. The number of activated genes $g$, the number of mRNA molecules $m$, and the number of protein molecules $p$ represent  $\mathcal G$. The time from last division itself is a stochastic variable for a heterogeneous population~\citep{harley1978cultured} and can be included as a parameter with the reaction rate constants. We assume that the conditional distribution $P(\mathcal G | \mathcal K)$ is known. Here, $\mathcal G = \{ g, m, p \}$ and $\mathcal K = \{ k_1, k_{-1}, k_2, k_3, k_4, k_5 \}$.

\subsection{The maximum entropy framework}

We now estimate the distribution of $\mathcal K$ using the maximum entropy (ME) framework~\citep{it_theory1}. A brief introduction to ME can be found in the supplementary material. Note that each point in the multidimensional $\mathcal K$-space represents a probability distribution in the $\mathcal G$-space. Consequently, the distribution whose entropy should be maximized is not $P(\mathcal K)$ but the joint distribution $P(\mathcal G, \mathcal K)$ of species and rates~\citep{caticha2004maximum,dixit2012}. 

The entropy  $S[P(\mathcal G, \mathcal K)]$ of the joint distribution $P(\mathcal G, \mathcal K)$ is given by
\begin{eqnarray}
S[P(\mathcal G, \mathcal K)] &=& -\sum_{\mathcal G,\mathcal K } P(\mathcal G,\mathcal K) \log P(\mathcal G, \mathcal K)\\
	&=& S[P(\mathcal K)] + \sum_{\mathcal K} S(\mathcal K) P(\mathcal K).
\end{eqnarray}
Here, \begin{eqnarray}P(\mathcal K) &=& \sum_{\mathcal G } P(\mathcal G,  \mathcal K), \\S[P(\mathcal K)] &=& -\sum_{\mathcal K } P( \mathcal K) \log P(\mathcal K)   \end{eqnarray}
and 
\begin{eqnarray}S(\mathcal K) &=& -\sum_{\mathcal G } P( \mathcal G | \mathcal K) \log P(\mathcal G | \mathcal K) \end{eqnarray} is the entropy of the conditional distribution $P(\mathcal G|\mathcal K)$.

If we constrain the mean values of the rate constants $\langle k_1 \rangle, \langle k_2 \rangle, \dots$, the ME framework predicts that the joint distribution maximizes the entropy $S[P(\mathcal G, \mathcal K)]$ subject to the constraints. To find the distribution, we introduce Lagrange multipliers $ \alpha_1, \alpha_2, \dots$ corresponding to rate constants $k_1, k_2, \dots$ and $\gamma$ for normalization. The modified objective function is
\begin{widetext}
\begin{eqnarray}
S[P(\mathcal G,\mathcal K)] &-& \sum_j \alpha_j \left ( \sum_{\mathcal G, \mathcal K}  P(\mathcal G, \mathcal K) k_j - \langle k_j \rangle\right ) + \gamma \left ( \sum_{\mathcal G,\mathcal K}  P(\mathcal G, \mathcal K)  - 1 \right ) \\
 =S[P(\mathcal K)] &+& \sum_{\mathcal K} S(\mathcal K)P(\mathcal K) - \sum_j \alpha_j \left ( \sum_{ \mathcal K}  P( \mathcal K) k_j - \langle k_j \rangle\right ) + \gamma \left ( \sum_{\mathcal K}  P(\mathcal K)  - 1 \right )  \label{eq:entr0}
\end{eqnarray}
\end{widetext}

Note that the mean values of the rate constants are not directly observable from experiments.  Employing them as constraints is a departure from the canonical understanding of the ME framework wherein probability distributions are predicted from moments calculated from experimental data. Yet, the ME framework can also be seen as an inference tool~\citep{shore,caticha2004maximum,dixit2012}: ME predicts the logically consistent probability distribution {\it if} mean values of certain important parameters of an experiment are fixed.

Since we know the functional form of $P(\mathcal G | \mathcal K)$, in Eq.~\ref{eq:entr0} we have summed over all possible values of $\mathcal G$ at a given value of $\mathcal K$. Setting the derivative of Eq.~\ref{eq:entr0} with respect to $P(\mathcal K)$ equal to zero and solving, we get,
\begin{eqnarray}
P(\mathcal K) &\propto& exp \left ( S(\mathcal K) - \sum_j \alpha_j k_j \right ) \label{eq:pk0}.
\end{eqnarray}
Eq.~\ref{eq:pk0} is the maximum entropy estimate of the distribution of  $\mathcal K$ if we constrain only the mean values of the rate constants. Notice that in addition to the usual exponentials (see supplementary materials), the distribution also depends on the entropy $S(\mathcal K)$ of the conditional distribution $P(\mathcal G|\mathcal K)$.

\subsection{Estimating $P(\mathcal K)$ in an $N$-reporter experiment}

Experimental advances allow us to construct more than one identical reporters for a gene inside a single cell~\citep{Elowitz2007,Stewart-Ornstein2012}. Mathematically, instead of generating samples of $\mathcal G$ from the distribution $P(\mathcal G| \mathcal K)$ for a fixed value of $\mathcal K$, we can conceive an experiment where we can sample $N$ identical experiments of the same species $\mathcal G$ from the joint distribution $P(\mathcal G_1, \mathcal G_2, \dots, \mathcal G_N|\mathcal K)$ at a fixed value of  $\mathcal K$. Note that the  variability in the extrinsic factors viz. the distribution $P(\mathcal K)$ bears no relation to  the number of reporters employed in a particular experiment. Consequently, we require the ME framework-predicted $P(\mathcal K)$ to be independent of  $N$~\cite{caticha2004maximum}.

If we assume that the $N$ experiments are sampled independently of each other --- this is a crucial assumption in $N$-reporter experiments~\citep{Elowitz2007,Stewart-Ornstein2012} --- we can write
\begin{eqnarray}
P(\mathcal G_1, \mathcal G_2, \dots, \mathcal G_N|\mathcal K) &=&\prod_{n=1}^{N} P(\mathcal G_n | \mathcal K ). \label{eq:ind}
\end{eqnarray}

Similar to the considerations above,  in order to estimate $P(\mathcal K)$ from this $N$-reporter experiment, we maximize the entropy of the joint distribution $P(\mathcal G_1, \mathcal G_2, \dots, \mathcal G_N, \mathcal K)$ constraining the mean values of the rate constants $\langle k_1\rangle, \langle k_2 \rangle, \dots$. The entropy of the joint distribution can be simplified using the independence in Eq.~\ref{eq:ind}
\begin{eqnarray}
S[P(\mathcal G_1, \mathcal G_2, \dots, \mathcal G_N, \mathcal K)] &=& S[P(\mathcal K)] + N\sum_{\mathcal K} S(\mathcal K)P(\mathcal K).
\end{eqnarray}

The modified objective function is given by (see Eq.~\ref{eq:entr0})
\begin{widetext}
\begin{eqnarray}
S[P(\mathcal K)] &+& N\sum_{\mathcal K} S(\mathcal K)P(\mathcal K) - \sum_j \alpha_j \left ( \sum_{ \mathcal K}  P( \mathcal K) k_j - \langle k_j \rangle\right ) + \gamma \left ( \sum_{\mathcal K}  P(\mathcal K)  - 1 \right )  \label{eq:entr1}
\end{eqnarray}
\end{widetext}

Consequently, the ME framework estimates the distribution $P(\mathcal K)$   as 
\begin{eqnarray}
P(\mathcal K) &\propto& exp \left( NS(\mathcal K) - \sum_j \alpha_j k_j\right ). \label{eq:pk1}
\end{eqnarray}

Interestingly, the estimate of the  variability $P(\mathcal K)$  depends on the number of reporters (see Eq.~\ref{eq:pk0} and Eq.~\ref{eq:pk1}) used in the experiment.  This problem will be aleviated if we introduce the average entropy of a given experiment $\langle S(\mathcal K) \rangle$ as an additional constraint. This additional constraint is not an experimentally observable constraint but merely a requirement of consistency in the prediction over multiple experiments~\citep{caticha2004maximum,dixit2012,Crooks2008}. Introducing the additional constraint $\langle S(\mathcal K) \rangle$ in the objective function by introducing a Lagrange multiplier $\mu_N$, we write the modified objective function as
\begin{widetext}
\begin{eqnarray}
S[P(\mathcal K)] &+& N\sum_{\mathcal K} S(\mathcal K)P(\mathcal K) - \sum_j \alpha_j \left ( \sum_{ \mathcal K}  P( \mathcal K) k_j - \langle k_j \rangle\right ) + \gamma \left ( \sum_{\mathcal K}  P(\mathcal K)  - 1 \right )   \nonumber \\&+& \mu_N \left ( \sum_{\mathcal K} S(\mathcal K)P(\mathcal K) -\langle S(\mathcal K) \rangle \right )\label{eq:entr2}
\end{eqnarray}
\end{widetext}
writing $N+\mu_N = \mu$ and maximizing with respect to $P(\mathcal K)$, we get
\begin{eqnarray}
P(\mathcal K) &\propto& exp \left( \mu S(\mathcal K) - \sum_j \alpha_j k_j\right ). \label{eq:pkmain}
\end{eqnarray}

Eq.~\ref{eq:pkmain} is the main theoretical result of this work. Briefly, if we know that the rate constants $\mathcal K$ vary from cell to cell and as a function of time and rather than precisely knowing them and if we constrain only their mean values, the ME framework predicts the distribution $P(\mathcal K)$  as Eq.~\ref{eq:pkmain}. Note that in addition to the usual exponentials, the distribution also depends on the conditional entropy $S(\mathcal K)$. Similar results have been obtained for thermodynamic systems~\citep{dixit2012,Crooks2008} and in estimating prior distributions in Bayesian inference~\citep{caticha2004maximum}.

\subsection{Experimentally observed distribution of chemical species}

The experimentally observable distribution $P(\mathcal G)$ is obtained by summing over all possible variations in $\mathcal K$. We get
\begin{eqnarray}
	P(\mathcal G; \mu, \alpha_1,\alpha_2,\dots) &\propto& \sum_{\mathcal K} P(\mathcal G|\mathcal K) \cdot exp \left( \mu S(\mathcal K) - \sum_j \alpha_j k_j\right ). \label{eq:pG}
\end{eqnarray}
Notice that the distribution in Eq.~\ref{eq:pG} is parametrized by $\mu$ and $\alpha_1, \alpha_2, \dots$. Each $\alpha_i$ corresponds to one rate constant $k_i$ while $\mu$ governs the extrinsic variability. In short, the ME framework predicts extrinsic variability only with one additional parameter $\mu$. Below, we will work out in detail the noise in the production of mRNA molecules from a constitutive promoter.

\subsection{The distribution of mRNA copy numbers}
Consider the simplified reaction scheme
\begin{eqnarray} 
{\rm DNA} \xrightarrow{\gamma} {\rm mRNA} \xrightarrow{\delta} \phi\end{eqnarray}
of transcription and degradation of mRNA molecules of a particular gene. $\gamma$ is the rate of transcription and $\delta$ is the rate of degradation. We have neglected the activation states of the DNA molecule e.g. promoter fluctuations~\citep{Kaufmann2007,Raj2006,Raj2008}. Promoter fluctuations are thought to occur, among other things, due to chromatin remodeling~\cite{golding2005real,Kaern2005,so2011general}. The chromosome of the DNA of a bacteria like {\it E. coli} is structured in $\sim 100-500$ nucleoids~\citep{reyes2008escherichia}. It is very likely that the chromatin structure  extends locally to $10-50$ genes around the gene studied and affects the transcription of {\it all} genes in a local region. Consequently, in a hypothetical dual-promoter experiment to study noise in mRNA production similar to~\citep{Elowitz2007}, promoter fluctuations are likely to affect the expression of {\it all} genes localized in a given region on the DNA in a correlated fashion. In what follows, we effectively treat promoter fluctuations as one of the local albeit extrinsic contributor to the variation in the {\it effective} rate of synthesis for the given gene. Below, we briefly discuss how to further parse the variability in the {\it effective} rate of synthesis into a contribution from promoter fluctuations and a contribution from other global extrinsic factors.

The solution of the reaction scheme at any time $t$ and at steady state is a Poisson distribution \begin{eqnarray}P({\rm m}|k)  = \frac{e^{-k}k^{{\rm  m}}}{{\rm m}!}\end{eqnarray} of mRNA copy number m with {\it effective} synthesis rate $k=\gamma/\delta \left (1 - e^{-\delta t} \right )$~\citep{Hemberg2007}.

The {\it effective} synthesis rate $k$ depends in a complicated manner on various factors including chromatin remodeling~\citep{golding2005real,so2011general,Kaern2005}, the states of many molecules in the cell including the components of RNA polymerase, the dynamics of assembly of the RNA polymerase holoenzyme,  various RNAse molecules, and other competing genes~\citep{Swain2002,Elowitz2007}. Consequently, it varies from cell to cell and also as a function of time from the start of the cell cycle. Thus, while studying gene expression in a population, instead of fixing a particular value of the effective synthesis rate $k$, we need to consider $P(k)$ the probability distribution of $k$. $P(k)$  quantifies the extrinsic contribution noisy gene expression.

For a given gene, experimentally assessing the  variability in $k$ is non-trivial --- $P(k)$ has to be {\it inferred} from limited experimental information viz. mean expression level, variation in gene expression level, etc. From Eq.~\ref{eq:pkmain}, we see that the distribution $P(k)$ is given by
\begin{eqnarray}
P(k) &\propto& exp \left [ (\mu\alpha - 1) S(k) -\alpha k \right ]. 
\end{eqnarray}
Here, $S(k)$ is the entropy of the conditional distribution $P({\rm m}|k)$, a Poisson distribution. Unfortunately, $S(k)$ does not have a closed form but $S(k) \sim \log k$. Thus, 
\begin{eqnarray}
P(k;\mu,\alpha) &\propto& k^{\mu\alpha-1}e^{-\alpha k}.\label{eq:pk}
\end{eqnarray}

In Eq.~\ref{eq:pk}, $\mu$ is the mean expression level and $\alpha = \eta_{\rm I}/\eta_{\rm E}$ is the ratio of the intrinsic and the extrinsic noise. The joint distribution $P({\rm m},k)$ is then given by,
\begin{eqnarray}
        P({\rm m},k) &=& P({\rm m}| k) P(k) \propto  \frac{e^{-\alpha k}k^{m+ \mu \alpha-1}}{{\rm m}!}.\label{eq:joint}
\end{eqnarray}
The experimentally accessible histogram  $P({\rm m})$ is obtained by summing over all variations in $k$ i.e. summing over the variation in extrinsic factors,
\begin{eqnarray}
P({\rm m}) &=& \sum_k P({\rm m},  k ) \propto \sum_k \frac{e^{-\alpha k}k^{m+ \mu \alpha-1}}{{\rm m}!}.
\end{eqnarray}
We estimate $P({\rm m})$ to be the negative binomial distribution (the discrete version of the gamma distribution),
\begin{eqnarray}
P({\rm m}) &\propto&\frac{1}{(1+\alpha)^{{\rm m}}} \times \frac{\Gamma \left [ {\rm m} + \alpha \mu \right ]}{{\rm m}!}. \label{eq:pjfinal}
\end{eqnarray}

\subsection{Noise decomposition of experimental data}
We estimate the total noise $\eta_{\rm T}$ from Eq.~\ref{eq:pjfinal} (see supplementary materials for details),\begin{eqnarray}\eta_{\rm T} &=& \frac{1}{\mu} \left ( 1 + \frac{1}{\alpha} \right) =   \frac{1}{\mu}  \left (1+\frac{\eta_{\rm E}}{\eta_{\rm I}} \right ) \ge \frac{1}{\mu} \nonumber
\end{eqnarray}
and,\begin{eqnarray}
\eta_{\rm I} &=& \frac{1}{\mu}~~{\rm ,}~~\eta_{\rm E} = \eta_{\rm T}- \frac{1}{\mu}. \label{eq:noises}\end{eqnarray}

The {\it greater than Poisson} relationship between $\eta_{\rm T}$ and the mean mRNA copy number $\mu$ (see Eq.~\ref{eq:noises}) is sometimes attributed to non-Poissonion dynamics e.g. promoter fluctuations, chromatin remodeling, mRNA synthesis bursts etc.~\citep{golding2005real,so2011general,Kaufmann2007,Raj2008,Raj2006,Taniguchi2010}. These effects themselves are thought to arise from cell to cell and dynamic variability in chromatin state and the state of DNA binding molecules~\citep{golding2005real,so2011general,Kaern2005}. Additionally, we suggest that the cell to cell variation in other extrinsic factors~\citep{Elowitz2007,Swain2002} also contributes to the {\it greater than Poisson} relationship.


The ME framework predicts that Eq.~\ref{eq:pjfinal} and Eq.~\ref{eq:noises} completely determine the histogram of mRNA copy numbers from experimentally measured mean expression level $\mu$ and total noise $\eta_{\rm T}$. Moreover,  $\eta_{\rm T}$ is always greater than 1 and $\eta_{\rm I}$ and $\eta_{\rm E}$ can be estimated from the histogram alone.  Importantly, the framework estimates the hitherto elusive  effect of extrinsic factors on gene expression viz. the distribution $P(k)$ of the {\it effective} synthesis rate $k$. 

The joint distribution Eq.~\ref{eq:joint} also allows us to estimate potentially interesting moments, for example, we predict that the Pearson correlation coeffient 
\begin{eqnarray}
\rho_{{\rm m}k} &=& \frac{1}{\sqrt{1+\alpha}} =\sqrt{ \frac{\eta_{\rm E}}{\eta_{\rm T}} } \label{eq:mkpred}
\end{eqnarray}
between {\it effective} mRNA synthesis rate and the mRNA copy number is  the square root of the ratio of extrinsic and total noise. These are some of the falsifiable predictions of the development presented here.

\section{Results and discussions}

\subsection{Numerical validation of the ME-predicted distribution}
\begin{figure*}[h]
        \includegraphics[scale=0.15]{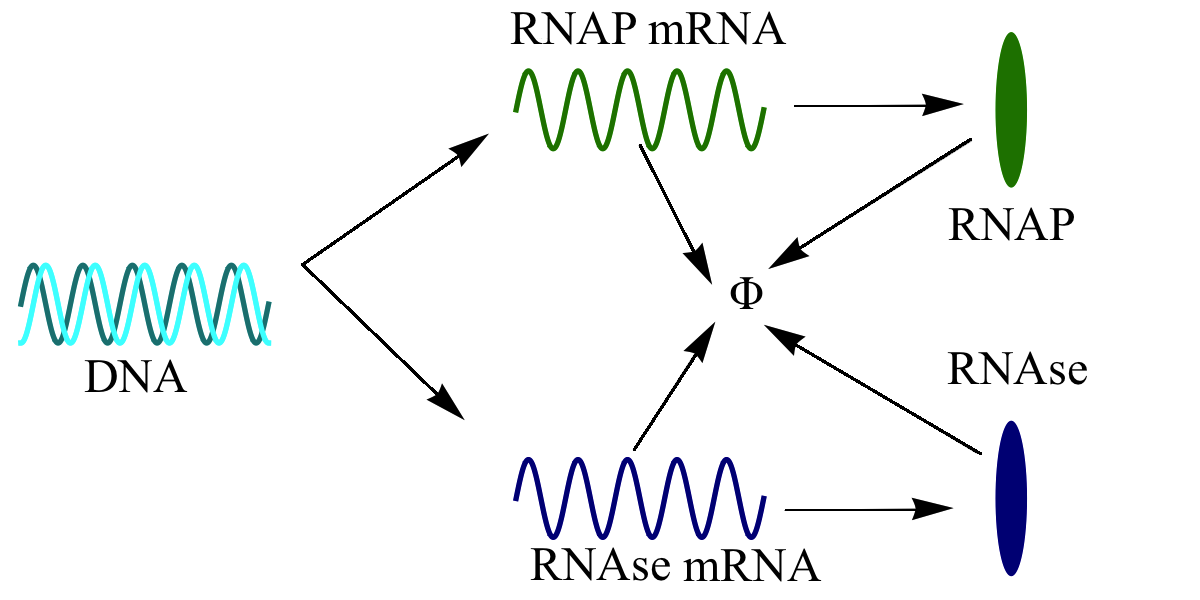}
        \includegraphics[scale=0.15]{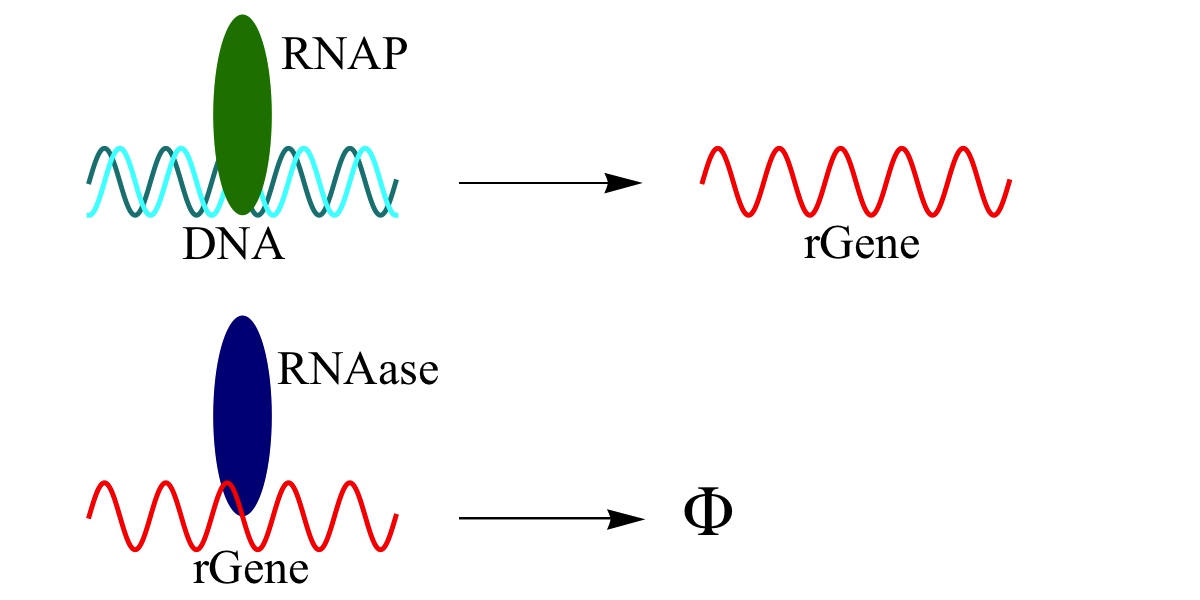}
        \caption{\label{fg:scheme}}
\end{figure*}

We analyze a simple numerical scheme for the synthesis of rGene, the mRNA of a constitutively expressed gene. In the scheme, the variability in the effective synthesis rate $k$ arises from the stochasticity in the production and degradation of the machinery (RNAP and RNAse). We show that the ME-predicted distribution  (Eq.~\ref{eq:pjfinal}) describes very accurately the numerically predicted distribution of  mRNA copy number for different strengths of extrinsic noise (see Fig.~\ref{fg:scheme} for a cartoon  and supplementary materials for details). 

\begin{figure*}[h]
        \includegraphics[scale=0.64]{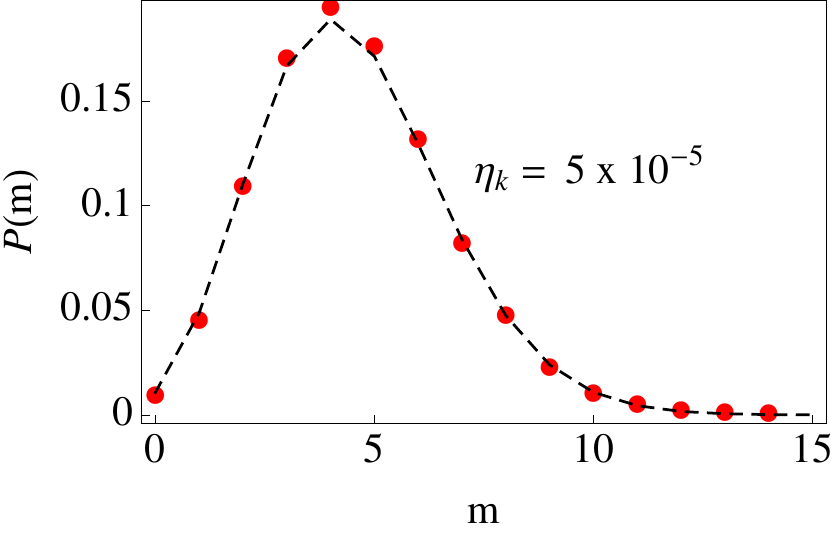}
        \includegraphics[scale=0.63]{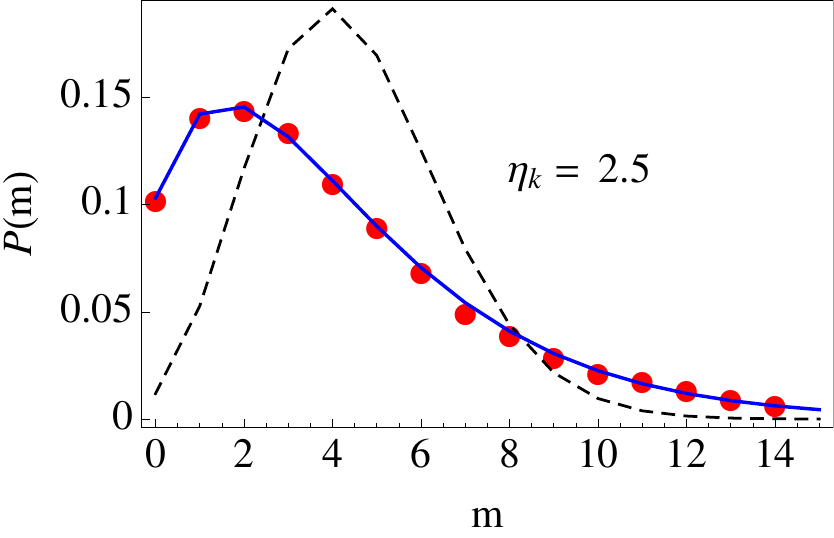}
        \includegraphics[scale=0.63]{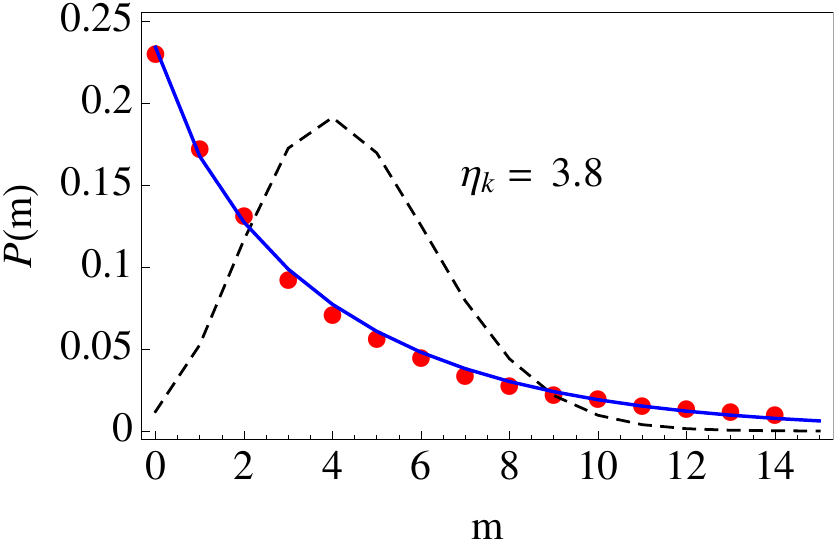}
        \caption{\label{fg:sim}}
\end{figure*}

Let [X] denote the concentration of species X.  In the model, the rate of synthesis $\gamma = \gamma_{0} [{\rm RNAP}]$ and the rate of degradation $\delta = \delta_{0}[{\rm RNAse}]$ of rGene, the mRNA of the gene under consideration, both depend on the concentration of the cellular proteins that carry out those reactions viz. [RNAP] (a proxy for the RNA polymerase complex) and [RNAse] (a proxy for RNAse) respectively.  Both the proteins are themselves are stochastically synthesized and degraded. The variation in the proxies mimics the cell to cell variations in extrinsic factors. The {\it effective} synthesis rate $k$ is directly proportional to the ratio [RNAP]/[RNAse]. We implement the Gillespie algorithm~\citep{gillespie} to estimate the steady state distribution of [rGene], the mRNA copy number. Even though the correlated dynamics of production of rGene, RNAP, and RNAse play an important part in determining the dynamics of the variability in [rGene], the steady state joint distribution [RNAP] and [RNAse] completely determines the steady state distribution of [rGene]. We only sample the distribution of mRNA copy numbers at long times ensuring that the steady state has been reached (see supplementary materials for details).
%
In order to clearly elucidate the effect of extrinsic factors on gene expression profile, in Fig.~\ref{fg:sim}, we show the  histogram of mRNA copy numbers for three different levels of noise, quantified by  \begin{eqnarray}
\eta_k &\equiv& \frac{\langle k^2 \rangle - \langle k \rangle^2}{\langle k \rangle^2} = \eta_{\rm E},
\end{eqnarray} the coefficient of variation in $k$, keeping the mean expression constant. The equality $\eta_k = \eta_{\rm E}$ is a consequence of the underlying single step process and will not hold true for other cases.

In the left panel of Fig.~\ref{fg:sim}, we show the histogram of mRNA copy numbers when the coefficient of variation $\eta_k$ is low ($\eta_k \approx 5 \times 10^{-5}$).  Observe that the histogram of mRNA copy numbers (red circles) is well described by a Poisson distribution (black dashes), as is expected. If we increase the variation in $k$ ($\eta_k \approx 2.5$ in the middle panel and $\eta_k \approx 3.8$ in the right panel),  the histogram of mRNA copy numbers gets broader and is best described by $P({\rm m})$  (Eq.~\ref{eq:pjfinal}, solid blue) rather than Poisson distribution (dashed black). Thus, even though the mRNA synthesis and degradation is governed by a Poisson process with an {\it effective} synthesis rate $k$, the  variation in the rate itself makes gene expression a doubly stochastic process~\citep{dsp,dixit2012} and leads to a histogram of mRNA copy numbers that is not Poisson-distributed and is best described by a Gamma-like distribution.

\subsection{Interpreting experiments}
\begin{figure}
        \includegraphics[scale=0.9]{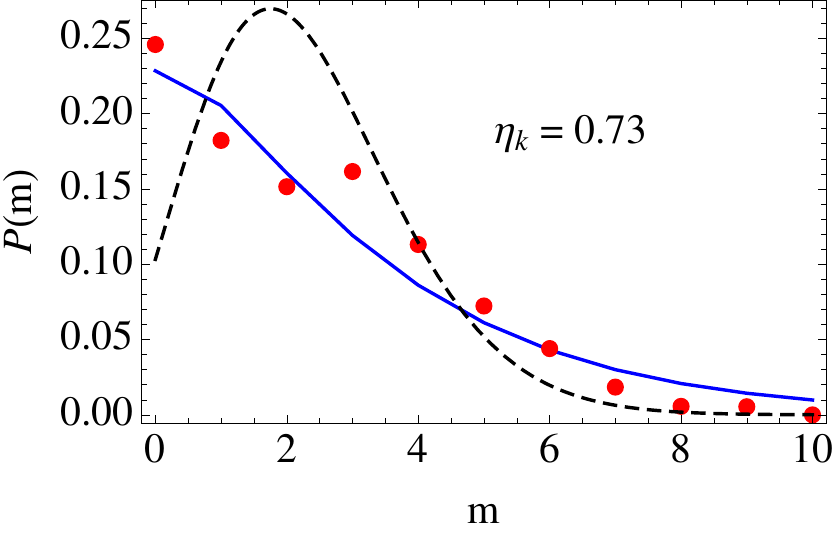}
        \caption{\label{fg:dist}}
\end{figure}

Fig.~\ref{fg:dist} shows the best fit to the histogram of mRNA copy numbers for the {\it E. coli} gene TufA~\citep{Taniguchi2010}. The Poisson distribution does not capture the mRNA histogram while Eq.~\ref{eq:pjfinal} describes it well (for a comparison with numerical simulations, see the right panel of Fig.~\ref{fg:sim}).  Also, recently,~\cite{so2011general} showed that the distribution of mRNA copy numbers in {\it E. coli} is well described by a Negative binomial distribution.
%

\begin{figure*}
        \includegraphics[scale=0.9]{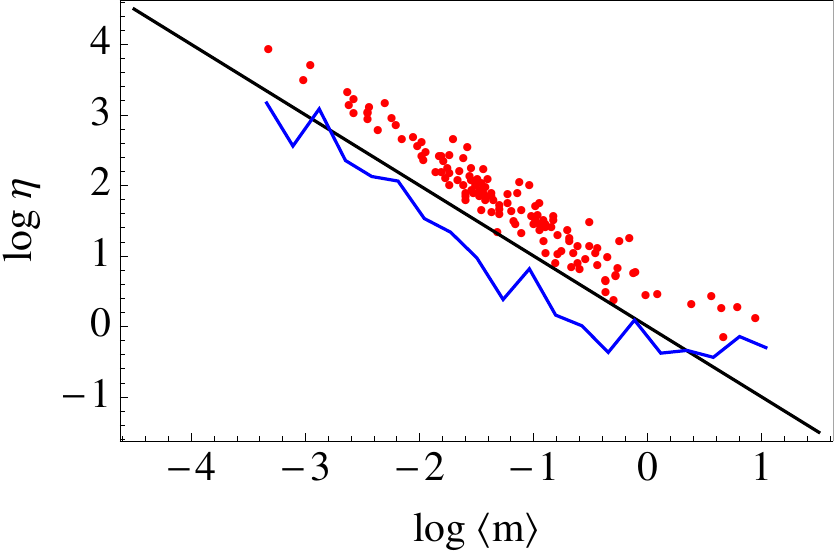}
        \caption{ \label{fg:noise}}
\end{figure*}

In Fig.~\ref{fg:noise}, we show the measured total noise and the predicted log-binned average trends in the decomposition of the total noise into its intrinsic and extrinsic components. The components are estimated from Eq.~\ref{eq:noises} for $\sim 130$ genes reported in~\cite{Taniguchi2010}. The noise decreases as mean expression level increases and both intrinsic and extrinsic components contribute significantly to the total noise. The total noise and the extrinsic noise saturate at high expression levels sometimes referred to as the `extrinsic limit'~\citep{Kaufmann2007,Taniguchi2010,Newman2006,Stewart-Ornstein2012}. Importantly, our framework also allows us to directly estimate the variation $P(k)$ of the {\it effective} synthesis rate $k$. 

\subsection{Incorporating promoter fluctuations explicitly}

The mRNA histogram from a slightly involved model that captures the activation state of the DNA molecule~\citep{Raj2006,golding2005real,so2011general} results in a distribution identical to Eq.~\ref{eq:pjfinal}. In that model, the deviation from Poisson distribtion is ascribed entirely to promoter fluctuations. As mentioned above, promoter fluctuations arise, among other things, due to chromatin remodeling~\citep{Kaern2005,golding2005real} and  are likely to affect the local region around the given gene~\citep{reyes2008escherichia}. Within our framework, the variation in mRNA synthesis rate due to promoter fluctuations is treated as extrinsic and is automatically incorporated in the distribution of the {\it effective} synthesis rate. 

It is a straightforward exercise within the current framework to further separate the variability in $k$ that arises due to promoter fluctuations from the variability that arises due to other extrinsic factors. The presence of other extrinsic factors can be tested in a number of ways. For example,  if promoter fluctuations are the major contributor to the variation of {\it effective} synthesis rate, it can be shown that the experimentally estimated skewness 
\begin{eqnarray}\gamma_1 &=& \frac{\langle {\rm m}^3 \rangle - 3\langle {\rm m} \rangle\langle {\rm m}^2 \rangle + 2\langle {\rm m}\rangle^3}{(\langle {\rm m}^2\rangle - \langle {\rm m} \rangle^2)^{3/2}} \end{eqnarray} 
of the distribution of mRNA numbers will be roughly equal to twice the square root of the total noise $\eta_{\rm T}$. In the presence of other extrinsic noise, this relationship is somewhat modified  (see supplementary materials for details). 

If promoter fluctuations are explicitly modeled,  the distribution of mRNA copy numbers  is characterized by at least two parameters~\cite{so2011general,Raj2006}. The development presented here will add one additional parameter to characterize the extrinsic variability beyond promoter fluctuations. Thus, the resulting distribution will be characterized by three parameters. Analyzing the currently reported experimental measurements of total noise to predict extrinsic noise beyond promoter fluctuations will consequently be an overfit. Yet, we note that if experimental measurements reliably estimate the third moment of the mRNA distribution, the current framework will be able to parse the total noise into its extrinsic and intrinsic (which will include promoter fluctuations) contributions without the assistance of a `two color' experiment (see supplementary materials for details).

\section{Concluding remarks\label{sc:disc}}

Measurements of the cell to cell variation in protein numbers show that the extrinsic contributions play a dominant role~\citep{Elowitz2007}. Yet, much of the theoretical development in understanding noise in gene expression has focused on the effect of intrinsic contributors viz. statistical mechanical fluctuations in binding and diffusion of molecules. The limited treatment extrinsic noise has received~\citep{Swain2002,Taniguchi2010,kaern:chaos2006} employs the linear fluctuation-dissipation like susceptibility analysis~\citep{Swain2002} or {\it ad hoc} assumptions about the nature of variation in extrinsic parameters~\citep{kaern:chaos2006,Taniguchi2010}.

To the best of our knowledge, we have, for the first time, presented a framework that systematically separates the intrinsic and the extrinsic contributors to noisy gene expression from limited information about the gene expression profile. We conclude that extrinsic factors can quantitatively and qualitatively change the experimentally accessible histogram of mRNA copy numbers. More importantly, the framework allows us to directly estimate the hitherto elusive variation in global extrinsic factors.

Specifically, we show that even if mRNA synthesis and degradation is described by a simple Poisson process, owing to the  variation in the {\it effective} synthesis rate $k$, the experimentally accessible histogram of mRNA copy numbers is broader and we estimate it to be the negative binomial distribution (see Eq.~\ref{eq:pjfinal}).  Consequently, we find that variation in the {\it effective} synthesis rate $k$ contributes to the  {\it greater than Poisson} relationship between noise $\eta_{\rm T}$ and the mean mRNA copy number $\langle {\rm m} \rangle$ (see Eq.~\ref{eq:noises}).  We also predict that in contrast to proteins~\citep{Elowitz2007} the variation in intrinsic and extrinsic factors both contribute significantly to the noisy expression of mRNA. Moreover, we directly probe the variation in {\it effective} mRNA synthesis rate $k$ and show that the coefficient of variation $\eta_k$ saturates at high expression levels (see bottom Fig.~\ref{fg:noise}).


Arguably, biologically interesting situations where noise is important are not limited to production of mRNA molecules. One would like to know how noise affects the regulation of internal circuits, response to external stimuli, and finally fitness and evolution. It is clear that once the distribution of  $\mathcal G$ is known as a function of  $\mathcal K$, the application of the current framework is in principle straightforward. Unfortunately, the conditional distribution $P(\mathcal G|\mathcal K)$ is known  for very few  simple cases (similar to the one discussed in this work). We propose the following algorithm to overcome this difficulty.

Even though the entire distribution $P(\mathcal G|\mathcal K)$  is almost always analytically inaccessible, the first two moments $\{ \langle \mathcal G_i \rangle_{\mathcal K} \} $ and $\{ \langle \mathcal G_i\mathcal G_j \rangle_{\mathcal K} \} $  can be estimated very  accurately  as analytical functions of  $\mathcal K$ for a number of complicated situation using the well known $\Omega$ expansion~\citep{Paulsson2005}. Moreover, under the assumption of linear noise, the entropy $S(\mathcal K)$ can itself be approximated as $S(\mathcal K) \sim \log {\rm det} \Sigma$ where $\Sigma_{ij} = \langle \mathcal G_i\mathcal G_j \rangle -\langle \mathcal G_i\rangle\langle \mathcal G_j\rangle$ is the covariance matrix. From here onwards, it is a straightforward exercise to compute $P(\mathcal K)$ using Eq.~\ref{eq:pkmain}. The intrinsic and extrinsic components can then be separated out analytically. We will implement the proposed program for protein synthesis and networks in the future.

\section{Acknowledgment}
I would like to thank Dr. Sergei Maslov, Dr. Adam de Graff for a critical reading and constructive suggestions. I would also like to thank Prof. Ken Dill, Prof. Dilip Asthagiri, and Ms. Shreya Saxena for stimulating conversations and suggestions about the manuscript.

This work was supported by grants PM-031 from the Office of Biological Research of the U.S. Department of Energy.

\bibliographystyle{biophysj}
\bibliography{gene}

\newpage

\section{Figure captions}

\begin{enumerate}
	\item The most general case of a constitutively expressing promoter. An inactive gene (black) is  turned into an active gene (and vice versa). The active gene (blue and green) is transcribed into an mRNA (red) which is then translated to a protein (red ellipse). The mRNA and the protein are also degraded. Various rate constants $\mathcal K$ govern the time evolution of $P(g,m,p| \mathcal K)$, the joint probability distribution of $g$ (number of activated genes), $m$ (number of mRNA molecules), and $p$ (number of protein molecules) is parametrized by the rate constants $\mathcal K$.
	\item A cartoon of the simplified scheme of mRNA production that takes into account extrinsic factors in gene expression levels (see supplementary materials for details). In the scheme, RNAP serves as the proxy for the RNA polymerase holoenzyme complex, RNAse is the proxy for RNA degradation machinery. The rate of synthesis of rGene, the RNA of a given gene is directly proportional to the concentration [RNAP] of the protein product of the RNAP gene. Similarly, the rate of degradation of rGene is directly proportional to the concentration [RNAse], the protein product of RNAse gene. RNAP and RNAse themselves are synthesized and degraded stochastically.
	\item The histogram of mRNA copy numbers (red dots), the Poisson distribution fit (black dashes) and the marginal distribution fit (solid blue, see Eq.~\ref{eq:pjfinal}) for three different scenarios in the numerical simulation. The mean mRNA copy number $\mu \approx 4.4$ for all three cases. Left: small variations in extrinsic factors ($\eta_k \approx 5 \times 10^{-5}$) results in a histogram of mRNA copy numbers that is well described by a Poisson distribution.  Middle: higher variation in extrinsic factors ($\eta_k \approx 2.5$) broadens the histogram of mRNA copy numbers. The marginal distribution $P({\rm m})$ (see Eq.~\ref{eq:pjfinal}) fits the data well. Right: high variation in extrinsic factors ($\eta_k \approx 3.8$) . Again, note that the histogram of mRNA copy numbers is wider than a Poisson distribution and the marginal distribution $P({\rm m})$ fits the simulation well.
	\item The observed histogram of mRNA copy numbers for the gene TufA~\citep{Taniguchi2010} is wider than a Poisson distribution (dashed black). Our results predict that the experimentally measured mRNA copy number histogram is described by Eq.~\ref{eq:pjfinal} (solid blue). $\eta_k \approx 0.7$ is the estimated coefficient of variation of the {\it effective} synthesis rate $k$.
	\item The experimentally measured total noise $\eta_T$ (red dots) is always higher than what is expected from a Poisson distribution (black line, see Eq.~\ref{eq:noises}). Our framework also allows us to predict the extrinsic noise $\eta_{\rm E}$ and the variation in the {\it effective} synthesis rate $\eta_k$. The blue line is the log-binned average of $\eta_{\rm E}$ (also equal to $\eta_k$). Notice that as opposed to proteins, for most mRNAs, intrinsic noise dominates the total noise for mRNAs. At higher mRNA numbers, the $\eta_{\rm E}$ dominates $\eta_{\rm T}$. Within the ME framework, we can explicitly estimate the hitherto inaccessible variation in the {\it effective} synthesis rate as well. 
\end{enumerate}

\pagebreak

\section{Figures}


\pagebreak

\newpage

\vspace{10mm}

\centerline{Supplementary materials}
\section{Maximum entropy formalism}
The maximum entropy formalism allows one to estimate the probability distribution $\{ p_i \}$ of states $\{i\}$ of a system from limited information. Briefly, the ME framework for estimating probabilities $\{p_i\}$ involves maximizing the entropy function $\mathcal S(\{p_i\})$ subject to constraining the values of certain  variables~\cite{it_theory1}. For example, if $\langle X_1 \rangle,\langle X_2 \rangle, \dots, \langle X_N \rangle$ are the mean values of variables $X_1, X_2, \dots, X_N$ respectively, then the probabilities $\{p_i\}$ of states $i$ are estimated by the maximizing the constrained objective function in Eq.~\ref{eq:maxEnt},
\begin{eqnarray}
\mathcal S (\{ p_i \}) &-& \sum_k  \lambda_k \left (  \left ( \sum_i p_i \cdot X_k(i)  \right ) - \langle X_k  \rangle \right )+ \alpha \left( \sum_i p_i - 1\right )  \label{eq:maxEnt}
\end{eqnarray}

Here, $\{ \lambda_k \}$ and $\alpha$ are Lagrange multipliers that ensure that the  constraints are satisfied and that the probabilities are normalized. The entropy is a non-negative convex function of the probabilities and is usually defined as~\cite{shore,shannon}
\begin{eqnarray}
\mathcal S(\{ p_i \}) &=& - \sum_i p_i \log p_i.
\end{eqnarray}

The maximization of Eq.~\ref{eq:maxEnt} estimates probabilities 
\begin{eqnarray}
p_i &=& \frac{1}{\mathcal Z(\{ \lambda_k \})} exp \left (- \sum_k \lambda_k X_k(i)\right ).
\end{eqnarray}
Here, 
\begin{eqnarray}
\mathcal Z(\{\lambda_k\}) &=& \sum_i exp \left (- \sum_k \lambda_k X_k(i)\right )
\end{eqnarray} is the partition function. The Lagrange multipliers $\lambda_k$ are determined by solving
\begin{eqnarray}
-\frac{\partial \log \mathcal Z}{\partial \lambda_k} &=& \langle X_k \rangle.
\end{eqnarray}

Notice that the probabilities depend exponentially on the constrained quantities (compare to Eq. 9, Eq. 13, and Eq. 15 in main text).

\section{Calculation of various moments}

If $P(k)$, $P({\rm m},k)$, and $P({\rm m})$ are given by  Eq. 20, Eq. 21, and Eq. 23, in the main text, the various moments are,
\begin{eqnarray}
\bar {\rm m} (k) &=&\sum_m mP({\rm m}|k) =  k,\\
\bar {\rm m}^2(k) &=& \sum_m m^2P({\rm m}|k) = k^2 + k,\\
\langle {\rm m} \rangle &=& \int \bar m(k) P(k) =  \langle k \rangle = \mu.
\end{eqnarray}
Similarly,
\begin{eqnarray}
\langle {\rm m}^2 \rangle &=&   \mu^2+ \mu + \frac{\mu}{\alpha} = \langle k^2 \rangle + \langle k \rangle,\\
\langle {\rm m} k\rangle &=& \mu^2 + \frac{\mu}{\alpha}.
\end{eqnarray}

The total noise is defined as
\begin{eqnarray}
\eta_{\rm T} &=& \frac{\langle {\rm m}^2 \rangle - \langle {\rm m} \rangle^2}{\langle {\rm m} \rangle^2} = \frac{1}{\mu} \left ( 1 + \frac{1}{\alpha}\right )
\end{eqnarray}

The intrinsic noise is defined as
\begin{eqnarray}
\eta_{\rm I} &=& \frac{1}{\langle {\rm m}\rangle^2} \int (\bar {\rm m}^2(k) - \bar {\rm m}(k)^2) P(k) dk =  \frac{1}{\langle {\rm m}\rangle^2}  \int k P(k) dk = \frac{1}{\mu}
\end{eqnarray}

\section{How to incorporate promoter fluctuations?}

When the promoter fluctuations are explicitly modeled, the distribution of mRNA copy numbers can be obtained in a closed form. Under simplifying conditions, the distribution of mRNA copy numbers becomes a negative binomial distribution~\cite{Raj2006,so2011general}. Here, we sketch a rough outline of incorporating extrinsic noise beyond promoter fluctuations within the ME framework.

For simplicity, let us assume that the mRNA copy number distribution $P({\rm m}; \alpha, \beta)$ is given by the Gamma distribution (the continuous counterpart of the negative binomial distribution). If there is no extrinsic noise beyond the promoter fluctuations, it is easy to show that 
\begin{eqnarray}
\langle {\rm m} \rangle &=& \alpha \beta,\\
\langle {\rm m}^2 \rangle &=& \alpha \beta^2 + \langle {\rm m} \rangle ^2,\\
\langle {\rm m}^3 \rangle &=& \alpha  (\alpha +1) (\alpha +2) \beta ^3.
\end{eqnarray}

The Gamma distribution has only two free parameters and the skewness is not independent of the second moment and is given by
\begin{eqnarray}
\gamma_1 &=& \frac{2}{\sqrt{\alpha}} = 2\sqrt{\eta_{\rm T}}. \label{eq:skew}
\end{eqnarray}
Eq.~\ref{eq:skew} roughly holds when promoter fluctuations are the major contributor to extrinsic noise. A deviation from Eq.~\ref{eq:skew} should prompt an exploration of the cell-to-cell variation in the parameters of the Gamma distribution themselves.

In real cells, the parameters $\alpha$ and $\beta$ of the Gamma distribution may be variable. The ME framework estimates the joint distribution $P(\alpha, \beta)$ from Eq. 15 of the main text. Even though the entropy of the Gamma distribution has a closed form, inserting the entropy in Eq. 15 of the main text and constraining the average value of $\alpha$ and $\beta$ results in an expression for $P(\alpha, \beta)$ that does not have a closed form. Instead, if we assume that $S(\alpha, \beta) \sim \log \sigma_{\rm m}$ i.e. the entropy scales as the variance of the mRNA copy number m (which is a good approximation), we get
\begin{eqnarray}
P(\alpha, \beta) &=& \frac{\zeta ^{\lambda +1} \xi ^{2 \lambda +1} \left(\alpha  \beta
   ^2\right)^{\lambda } e^{-\alpha  \zeta -\beta  \xi }}{\Gamma (\lambda
   +1) \Gamma (2 \lambda +1)}. \label{eq:pab}
\end{eqnarray}
Eq.~\ref{eq:pab} is equivalent to Eq. 20 in the main text when promoter fluctuations are explicitly modeled. Notice that the distribution of the parameters $\alpha$ and $\beta$ themselves is described by a product of two independent Gamma distribution. The variability in $\alpha$ and $\beta$ can now be ascribed to other global extrinsic factors.

Notice that $P(\alpha, \beta)$ is parametrized by three parameters $\lambda$, $\zeta$, and $\xi$. The resuling marginal distribution $P({\rm m})$ for m will also be parametrized by three parameters. Unfortunately, this marginal distribution doesn't have a closed form either. Yet, we can indeed compute quantities such as the total, intrinsic, and extrinsic noise from Eq.~\ref{eq:pab}. Notice that since $P({\rm m})$ has three free parameters, the skewness estimated from Eq.~\ref{eq:pab} may not be equal to twice the square root of the total noise. Computing various moments from Eq.~\ref{eq:pab}, we get
\begin{eqnarray}
\langle {\rm m} \rangle &=& \frac{(\lambda +1) (2 \lambda +1)}{\zeta  \xi },\\
\langle {\rm m}^2\rangle &=& \frac{2 (\lambda +1)^2 (2 \lambda +1) (\zeta +\lambda +2)}{\zeta ^2 \xi
   ^2}, \\
\langle {\rm m}^3 \rangle &=& \frac{2 (\lambda +1)^2 (2 \lambda +1) (2 \lambda +3) \left(3 \zeta 
   \lambda +2 \zeta  (\zeta +3)+\lambda ^2+5 \lambda +6\right)}{\zeta ^3
   \xi ^3}
\end{eqnarray}

The intrinsic, extrinsic, and the total noise are given by,
\begin{eqnarray}
\eta_{\rm I} &=& \frac{2 \zeta }{2 \lambda +1},\\
\eta_{\rm E} &=& \frac{3}{2 \lambda +1},\\
\eta_{\rm T} &=& \frac{2 \zeta +3}{2 \lambda +1}.
\end{eqnarray}

And the skewness $\gamma_1$ is
\begin{eqnarray}
\gamma_1 &=& \frac{2 \sqrt{\lambda +1} \sqrt[4]{2 \lambda +1} \left(6 \zeta ^2+(4 \zeta  (\zeta +3)+11) \lambda +15 \zeta +13\right)}{(2 \zeta +3)^{3/4} (\zeta  \xi )^{3/2}}.
\end{eqnarray}

The developed framework can potentially parse intrinsic and extrinsic contributions if higher moments of the mRNA copy number are carefully estimated. Even though the theoretical framework allows it, unfortunately, currently published experimental data does not permit us to do the same.

\section{Numerical simulations}

The synthesis and degradation of the mRNA of any given gene competes with the synthesis and degradation of all other co-expressed genes. Moreover, the cellular machinery that carries out these reactions itself comprises of proteins and mRNAs and is subject to cell to cell variation. We devise a simple scheme to mimic the coupled dynamics of synthesis and degradation of the cellular machinery with the dynamics of synthesis and degradation of the mRNA of a given gene.
\begin{center}  
	\begin{table*}
		{\bf Parameters for the simulation} \\                
		\begin{tabular}{c|ccc}
                  Parameter & Case 1 & Case 2 & Case 3 \\ \hline
		$\gamma_1$ & 2.0& 2.0  &2.0 \\
		$\gamma_2$ & 2.0& 2.0  & 2.0\\
		$\gamma_0$ & 0.9 & 1.6  & 1.0 \\
		$\kappa_1$ & 0.5 & 0.5   &  0.5\\
		$\kappa_2$ & 0.5&  0.5  & 0.5\\
		$\delta_1$ & 0.1 & 0.1  & 0.1\\
		$\delta_2$ & 0.1 & 0.1 & 0.1\\
		$\delta_0$ & 0.227 & 0.5  & 0.1 \\
		$\Delta_1$ & 0.15 & 0.65 & 0.55\\
		$\Delta_2$ & 0.15 & 0.65 & 0.55 \\
		$[{\rm DNA}]_{\rm RNAP}$  & 5 &5 &5 \\
		$[{\rm DNA}]_{\rm RNAase}$  & 5 & 5&5 \\
		$[{\rm DNA}]_{\rm Gene}$  & 1 &1 & 1\\
  		\end{tabular}                 
		 \caption{The details of the parameters for the numerical simulation of mRNA synthesis. All rates are in s$^{-1}$ and all copy numbers are integers.\label{table:details}}
        \end{table*}
\end{center}

The transcription apparatus is represented by a single protein RNAP and the mRNA degradation apparatus
is represented by a single protein RNAase. The rate of synthesis $\gamma$ of the given mRNA depends linearly on [RNAP]  the concentration of the proxy for the RNA polymerase complex. Similarly, the rate of degradation $\delta$ depends linearly on the concentration [RNAase] of the proxy for the RNAase enzyme ($\gamma = \gamma_0 [{\rm RNAP}]$ and $\delta = \delta_0 [{\rm RNAase}]$).
\begin{eqnarray*}
{\rm DNA}& \xrightarrow{\gamma_1}& {\rm rRNAP} \\
{\rm DNA} &\xrightarrow{\gamma_2}& {\rm rRNAase} \\
{\rm DNA} &\xrightarrow{\gamma}& {\rm rGene} \\
{\rm rRNAP} &\xrightarrow{\kappa_1}& {\rm RNAP}\\
{\rm rRNAase} &\xrightarrow{\kappa_2}& {\rm RNAase}\\
{\rm rRNAP} &\xrightarrow{\delta_1}& {\rm \phi}\\
{\rm rRNAase} &\xrightarrow{\delta_2}& {\rm \phi}\\
{\rm rGene} &\xrightarrow{\delta}& {\rm \phi}\\
{\rm RNAP} &\xrightarrow{\Delta_1}& {\rm \phi}\\
{\rm RNAase} &\xrightarrow{\Delta_2}& {\rm \phi}\\
\end{eqnarray*}

The dynamics of the synthesis and degradation of the mRNA of the given gene and RNAP and RNAase is propagated using the Gillespie's algorithm~\cite{gillespie} for $2 \cdot 10^8$ steps. Data is stored every 5000th step after an initial equilibration of 50000 steps. The initial concentrations of all species except the copy number of the each gene on the DNA at $t=0$ was set to 0. Table~\ref{table:details} gives the details of the conditions that were employed to construct the histograms (red points in Fig. 2 of the main text).

\end{document}